\documentclass[aps,prc,showpacs]{revtex4}
\usepackage[dvips]{graphicx}
\usepackage[dvips]{color}
\usepackage{amsmath}
\usepackage{float}

\newcommand{\bea}{\begin{eqnarray}}
\newcommand{\eea}{\end{eqnarray}}
\newcommand{\be}{\begin{equation}}
\newcommand{\ee}{\end{equation}}

\newcommand{\sumint}{\sum\kern -3.5ex \int\kern 1.0ex}

\newlength\dlf  

\begin{document}

\title{Density dependence of 2p-2h meson-exchange currents
}

\author{
J.E. Amaro$^a$,
M.B. Barbaro$^{b,c}$,
J.A. Caballero$^d$,
A. De Pace$^c$,
T.W. Donnelly$^e$,
G.D. Megias$^d$, 
I. Ruiz Simo$^a$
}

\affiliation{$^a$Departamento de F\'{\i}sica At\'omica, Molecular y Nuclear,
and Instituto de F\'{\i}sica Te\'orica y Computacional Carlos I,
Universidad de Granada, Granada 18071, Spain}

\affiliation{$^b$Dipartimento di Fisica, Universit\`a di Torino,
 Via P. Giuria 1, 10125 Torino, Italy}

\affiliation{$^c$INFN, Sezione di Torino, Via P. Giuria 1, 10125 Torino, Italy}  
\affiliation{$^d$Departamento de F\'{\i}sica At\'omica, Molecular y Nuclear,
Universidad de Sevilla, Apdo.1065, 41080 Sevilla, Spain}

\affiliation{$^e$Center for Theoretical Physics, Laboratory for Nuclear
  Science and Department of Physics, Massachusetts Institute of Technology,
  Cambridge, MA 02139, USA}

\date{\today}


\begin{abstract}
  We analyze the density dependence of the contribution of
  meson-exchange currents to the lepton-nucleus inclusive cross
  section in the two-particle two-hole channel.  The model is based on
  the Relativistic Fermi Gas, where each nucleus is characterized by
  its Fermi momentum $k_F$. We find that the 2p-2h nuclear response
  functions at their peaks scale as $A k_F^2$ for Fermi momentum going from 200 to
  300 MeV/c and momentum transfer $q$ from $2k_F$ to 2 GeV/c. This
  behavior is different from what is found for the quasielastic response,
  which scales as $A/k_F$. Additionally, the deep scaling region is also discussed and there the usual scaling behavior is found to be preferable.
\end{abstract}

  

\maketitle

\section{Introduction}
\label{intro}

Two-particle two-hole (2p-2h) excitations in electroweak nuclear
reactions have been extensively explored in the past
\cite{Van81}-\cite{Ama10a} in electron and
neutrino scattering studies.  These states, where two nucleons are
promoted above the Fermi level leaving two holes inside the Fermi sea,
are known to give a large contribution to the inclusive $(e,e')$ cross
section in the so-called ``dip region'', corresponding to excitation
energies lying between the quasielastic (QE) and $\Delta(1232)$
excitation peaks.

This subject has received renewed attention in recent years, since
2p-2h excitations have been shown to play an important role in
explaining neutrino-nucleus cross sections measured in neutrino
oscillation experiments
\cite{Mini1,Mini2,Minerva1,Minerva2,T2K1,T2K2}.  Whereas most of the
existing calculations refer to a $^{12}$C target
\cite{Mar09,Ama10b,Nie11,Ben15,Sim16,Meg16e,Meg16nu},
 there is
  growing interest in the extension to heavier nuclei, such as
  $^{16}$O, $^{40}$Ar, $^{56}$Fe and $^{208}$Pb, used in ongoing and
  future neutrino experiments.  Since the calculation of the 2p-2h
  response is computationally demanding and time consuming, in this
  paper we
provide an estimate of the density dependence of
  these contributions which can be used to extrapolate the results
  from one nucleus to another.

\vspace{0.2cm}

In \cite{Don99-1,Don99-2} inclusive electron scattering data
from various nuclei were analyzed in terms of ``superscaling'': it
was shown that, for energy loss below the quasielastic peak, the scaling
functions, represented versus an appropriate dimensionless scaling
variable, are not only independent of the momentum transfer (scaling
of first kind), but they also coincide for mass number $A\geq$4 (scaling
of second kind).  More specifically, the reduced QE cross section
was found to scale as
$A/k_F$, $k_F$ being the Fermi momentum. The Fermi momenta typical of
most nuclei belong to the range 200--300 MeV/c ~\cite{Chiara}.
It was also shown that for higher energy transfers superscaling is
broken and that its violations reside in the transverse channel rather
than in the longitudinal one. Such violations must be ascribed to
reaction mechanisms different from one-nucleon
knockout. Two-particle-two-hole excitations, which are mainly transverse and
occur in the region between the quasielastic and $\Delta$ production
peaks, are -- at least in part -- responsible for this violation.

  In this paper we explore the $k_F$-dependence of the 2p-2h nuclear
  response evaluated within the model of \cite{DePace03}, based on
  the Relativistic Fermi Gas (RFG).  The model has recently been
  extended to the weak sector \cite{Sim16} and applied to the study of
  neutrino-nucleus scattering.  We refer the reader to the original
  papers for the details of the model. Here we just mention its main
  features: it is based on a fully relativistic Lagrangian including
  nucleons, pions and $\Delta$ degrees of freedom; it involves the
  exact calculation of a huge number of diagrams, each of them
  involving a 7-dimensional integral; and it takes into account
  both direct and exchange Goldstone diagrams.

\vspace{0.2cm}

\section{Formalism}  

The lepton-nucleus inclusive cross section can be described in terms
of response functions, which embody the nuclear dynamics. 
There are  two response
functions in the case of electron scattering,
\begin{equation}
\frac{d^2\sigma}{d\Omega
  d\omega}=\sigma_{Mott}\left[v_LR^L(q,\omega)+v_TR^T(q,\omega )
  \right] \, ,
\label{d2sem}
\end{equation}
and five in the case of charged-current (anti)neutrino scattering,
\begin{equation}
\frac{d\sigma}{dk' d\Omega}=\sigma_0\left[\hat V_{CC}R^{CC}(q,\omega)+
  2 \,\hat V_{CL}R^{CL}(q,\omega)+\hat V_{LL}R^{LL}(q,\omega)+\hat
  V_TR^T(q,\omega)\pm2\hat V_{T'}R^{T'}(q,\omega)\right] \, .
\label{d2scc}
\end{equation}
In the above $\sigma_{Mott}$ is the Mott cross section, $\sigma_0$ the
analogous quantity for neutrino scattering, $q$ and $\omega$ the
momentum and energy transferred to the nucleus, $\Omega$ and
$k^\prime$ the outgoing lepton solid angle and momentum, and $v_K$,
$\hat V_K$ kinematical factors that only depend on the leptonic
variables (see \cite{Amaro:2004bs} for their explicit
expressions). The $\pm$ sign in Eq. (\ref{d2scc}) refers to neutrino and
antineutrino scattering, respectively.  We shall denote by $R^K_{\rm
  MEC}$ the contribution to the response $R^K$ arising from the
excitation of 2p-2h states induced by meson-exchange currents (MEC).

In order to remove the single-nucleon physics from the problem (which
also causes the fast growth of the response as $\omega$ approaches the
light-cone), it is useful to define the following reduced response
(per nucleon) 
\be
F^T_{\rm MEC}(q,\omega) \equiv
\frac{R^T_{\rm MEC}(q,\omega)}{\widetilde G_M^2(\tau)} \,,
\label{eq:FT}
\ee where $\tau\equiv (q^2-\omega^2)/(4m_N^2)$ and
\be \widetilde G_M^2(\tau) \equiv Z G_{Mp}^2(\tau) + N G_{Mn}^2(\tau) \,, \ee
$G_{Mp}$ and $G_{Mn}$ being the proton and neutron magnetic form
factors. For simplicity here we neglect in the single-nucleon dividing 
factor small contributions coming from the motion of the nucleons,
  where the electric form factor contributes, which depend on
  the Fermi momentum \cite{scaling}.

Since the behavior with density of the nuclear response is not
expected to depend very much on the specific channel or on the nature
of the probe, for sake of illustration we focus on the electromagnetic
2p-2h transverse response, which largely dominates over the
longitudinal one. 
  Our starting point is therefore
the electromagnetic transverse response, $R^T_{\rm MEC}$, associated
with meson-exchange currents (MEC) carried by the pion and by the
$\Delta$-resonance, evaluated within the model of \cite{DePace03}.

\section{Results}
%
\begin{figure}[!htb]
\begin{center}
\includegraphics[scale=1.0]{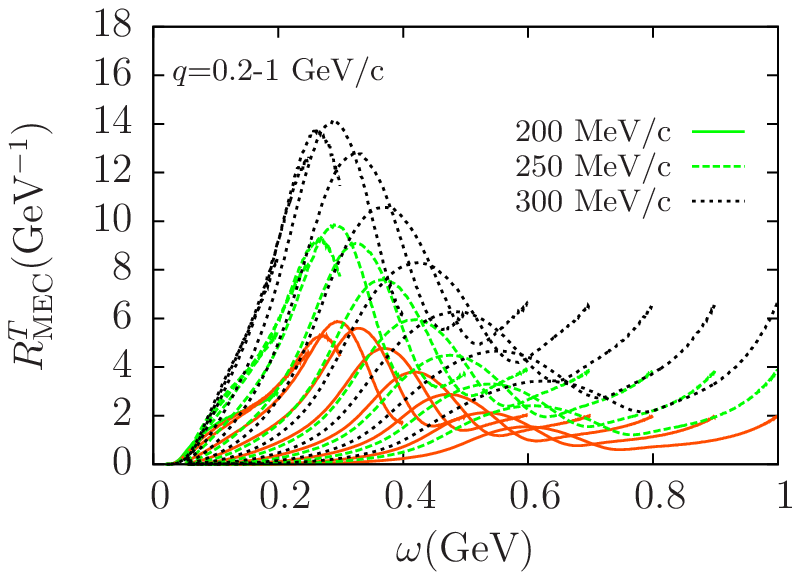}\hspace{0.5cm} \includegraphics[scale=1.0]{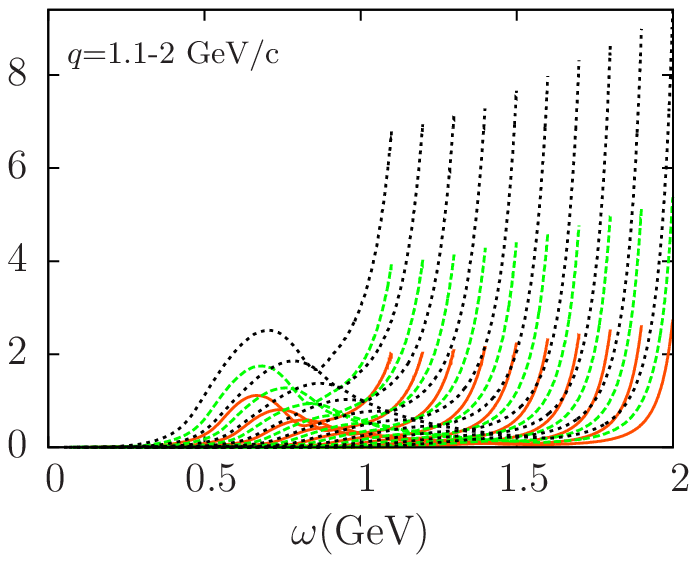}
\caption{(Color online) The 2p-2h MEC response of \cite{DePace03} plotted versus  $\omega$ for three values of the Fermi momentum $k_F$ and for different values of the momentum transfer $q =$ 200, \ldots, 1000 MeV/c (left panel) and 1100, \ldots 2000 MeV/c (right panel), increasing from left to right.}
\label{fig:fig1}
\end{center}
\end{figure}%
%

\begin{figure}[!htb]
\begin{center}
\includegraphics[scale=1.0]{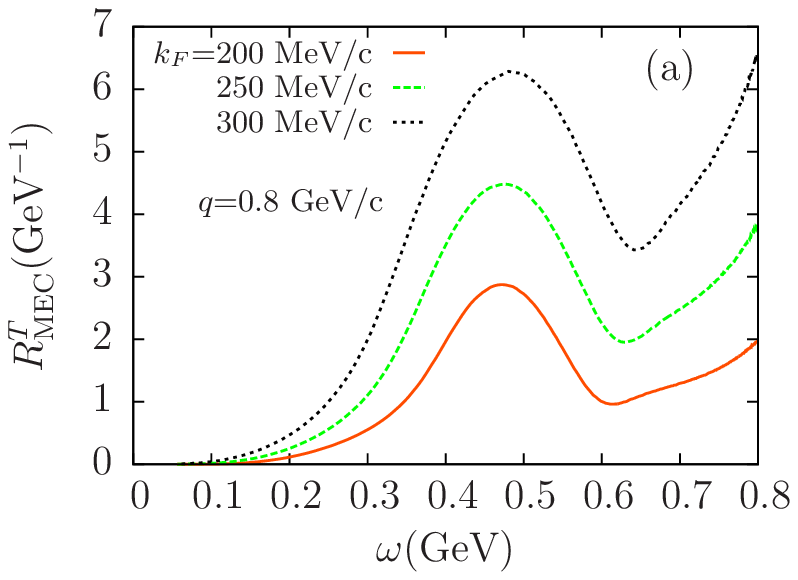}\hspace{0.5cm}\includegraphics[scale=1.0]{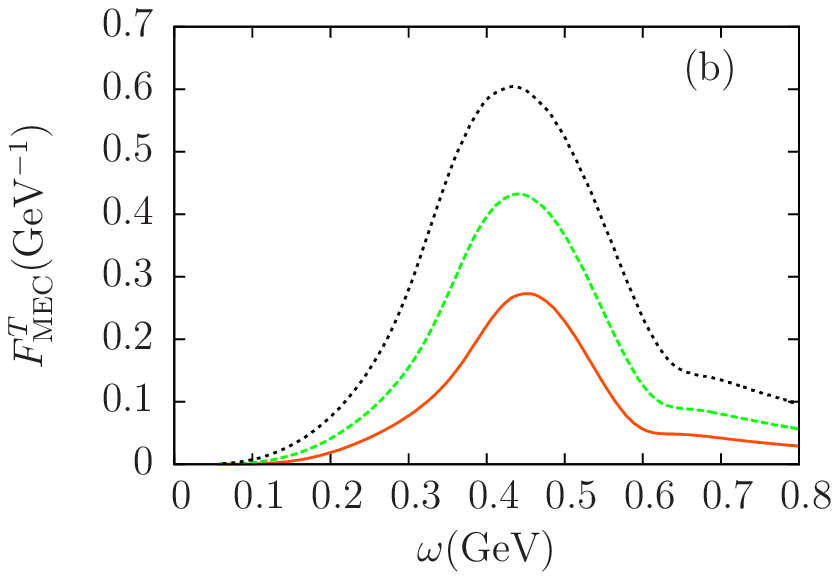}
\includegraphics[scale=1.0]{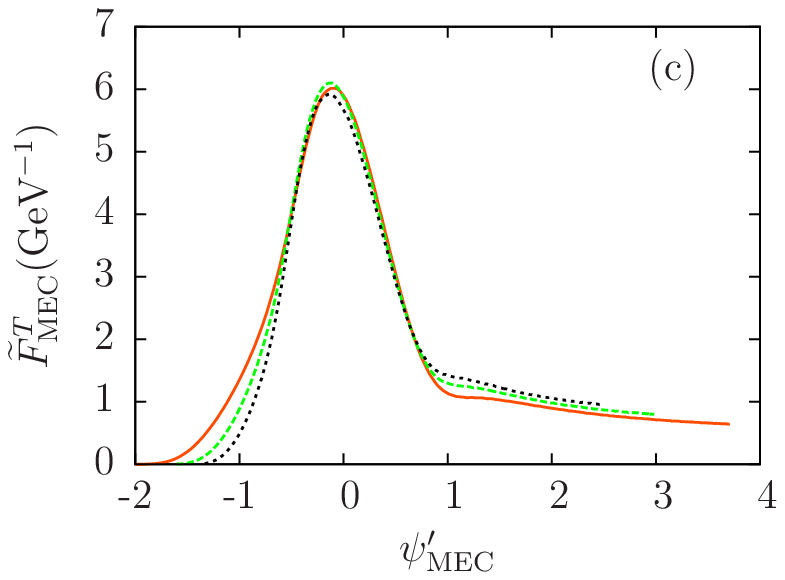}\hspace{0.5cm}\includegraphics[scale=1.0]{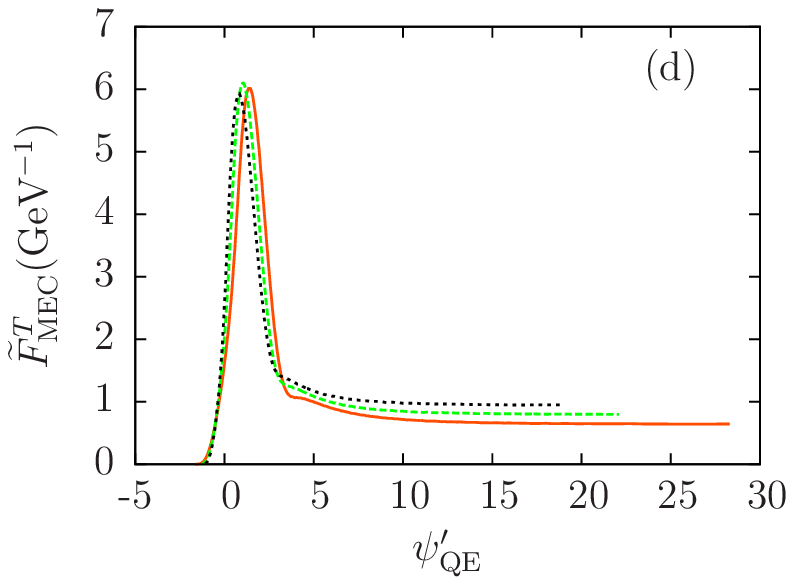}
 \caption{(Color online) Upper panels: the 2p-2h MEC response (a) and the reduced
   response defined by Eq. (\ref{eq:FT}) (b) plotted versus $\omega$
   for $q$=800 MeV/c and Fermi momentum $k_F$ varying between 200
   (lower curve) and 300 (upper curve) MeV/c.  Lower panels: the
   corresponding scaled 2p-2h MEC response defined by
   Eq. (\ref{eq:tilf}) plotted versus the scaling variables
   $\psi^\prime_{\rm MEC}$ (c) and $\psi^\prime_{\rm QE}$
   (d). }
\label{fig:fig2}
\end{center}
\end{figure}
%

%
\begin{figure}[!htb]
\begin{center}
\includegraphics[scale=0.9]{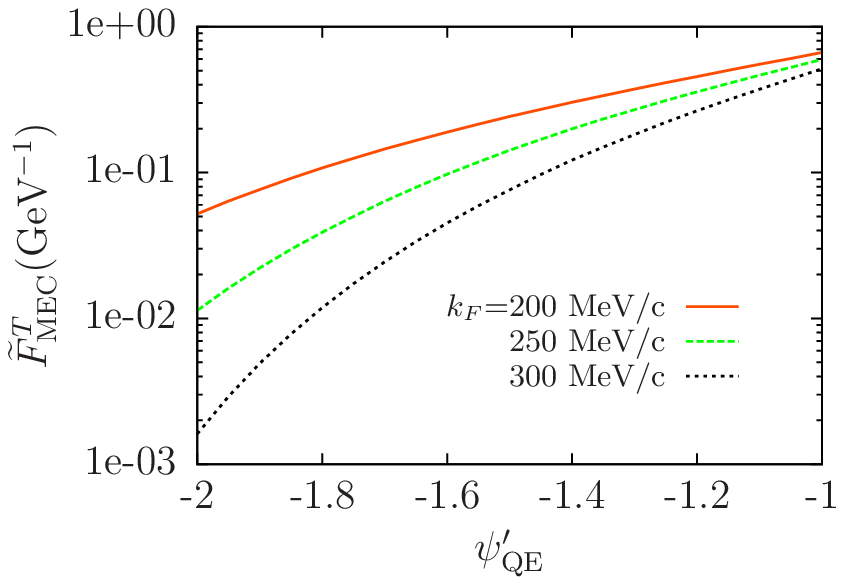}\hspace{0.5cm}\includegraphics[scale=0.9]{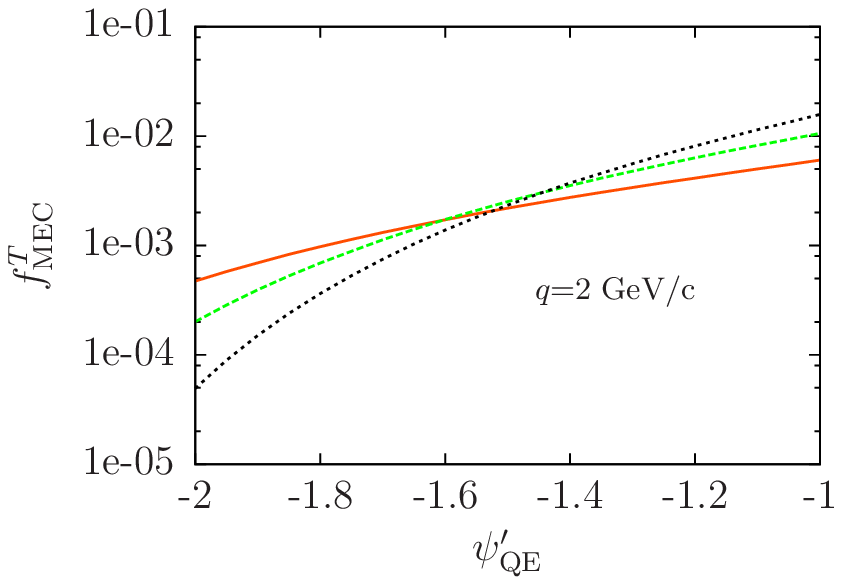}
\caption{ (Color online) The scaled 2p-2h MEC response defined by
  Eq. (\ref{eq:tilf}) (left panel) and the corresponding superscaling function defined by  
  Eq. (\ref{eq:fsup}) (right panel) plotted versus the scaling variable
  $\psi^\prime_{\rm QE}$ for $q$=2 GeV/c.}
\label{fig:fig3}
\end{center}
\end{figure}
%

\begin{figure}[!htb]
\begin{center}
\includegraphics[scale=1.0]{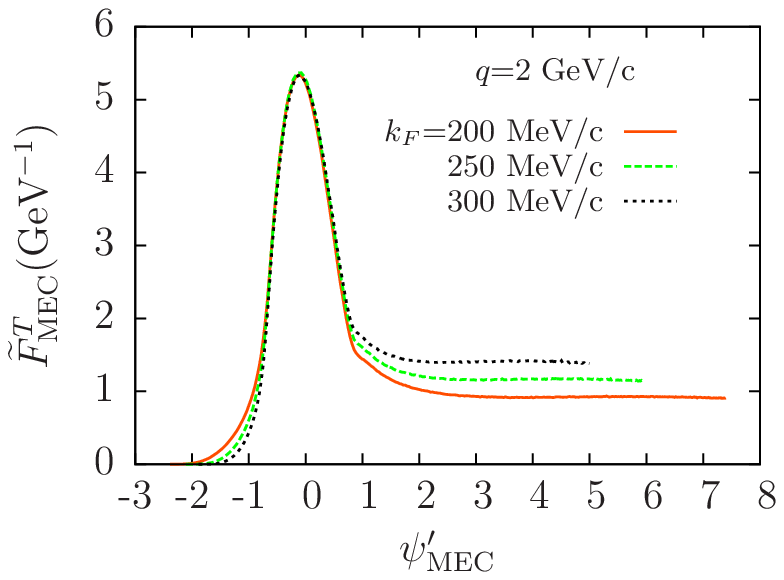}\hspace{0.5cm}\includegraphics[scale=1.0]{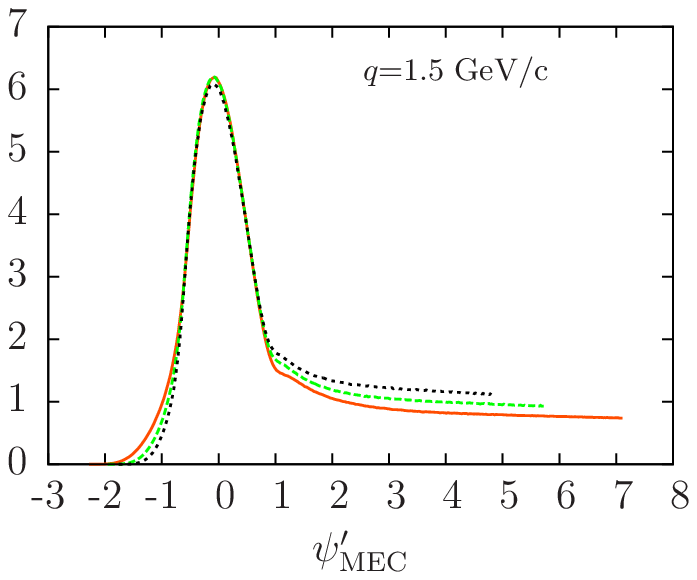}
\includegraphics[scale=1.0]{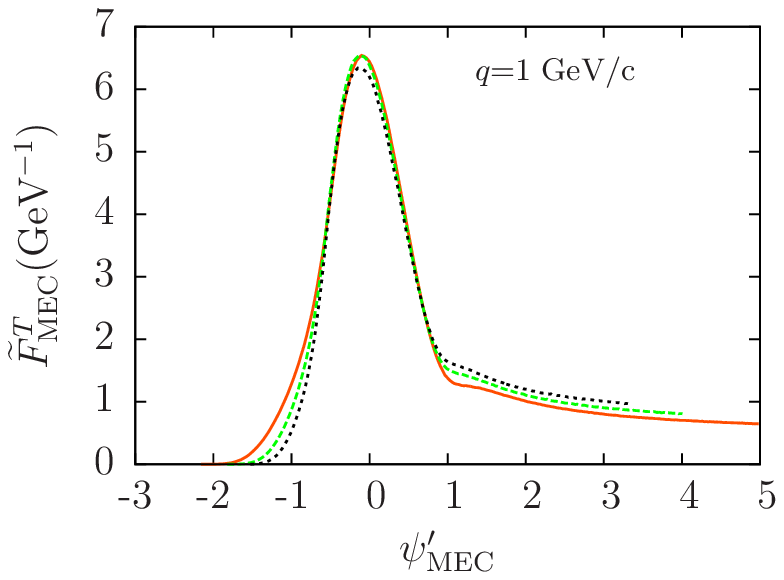}\hspace{0.5cm}\includegraphics[scale=1.0]{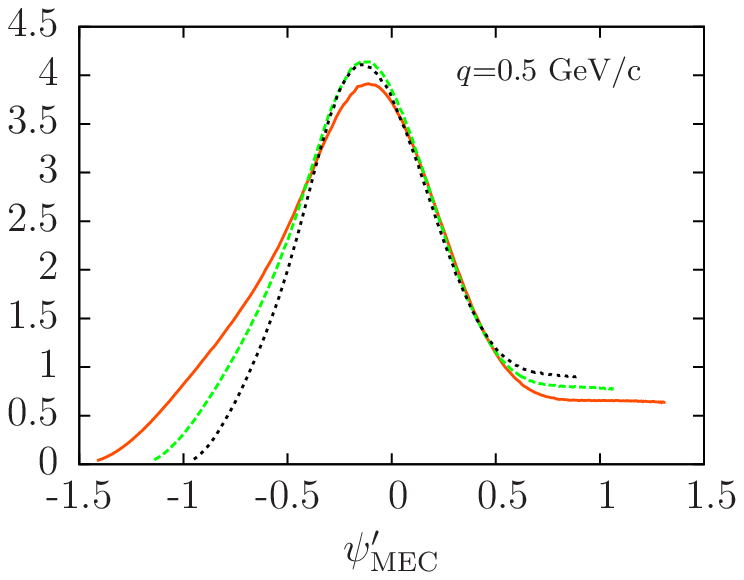}
\caption{(Color online) As for Fig. 2c, but now for different values of $q$.}
\label{fig:fig4}
\end{center}
\end{figure}
%

\begin{figure}[!htb]
\begin{center}
\includegraphics[scale=1.0]{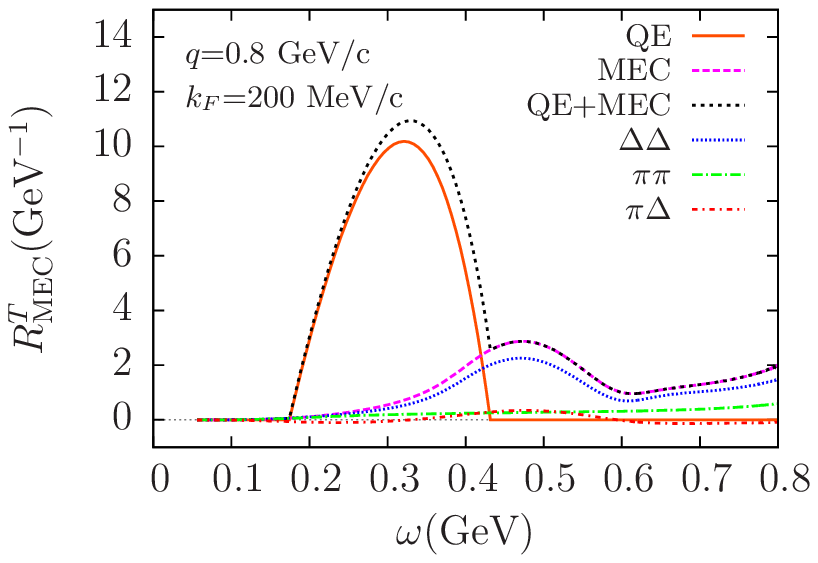}\hspace{0.5cm}\includegraphics[scale=1.0]{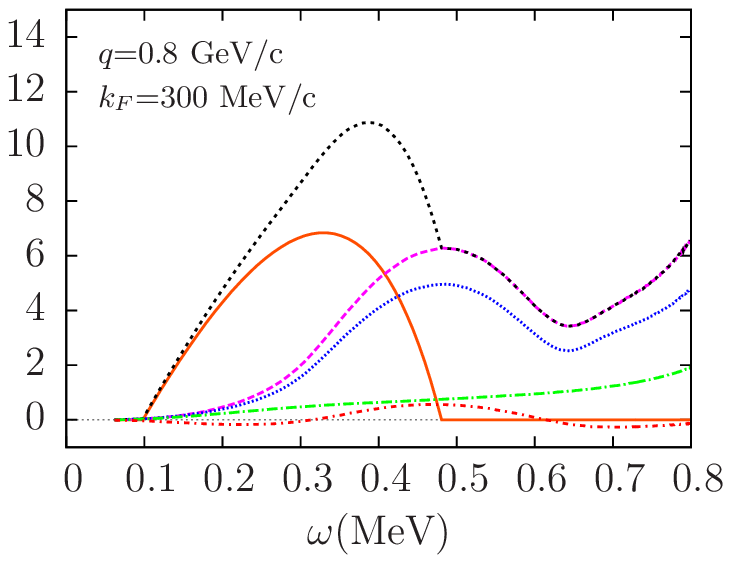}
\includegraphics[scale=1.0]{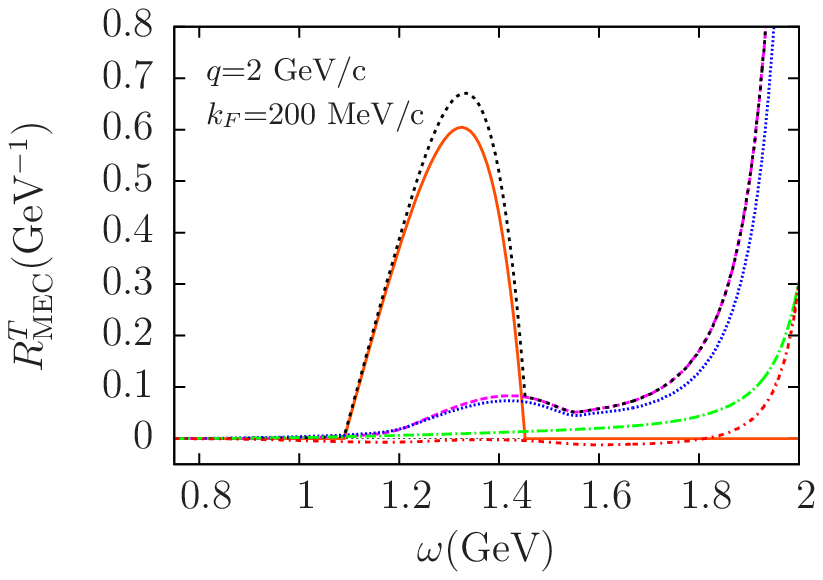}\hspace{0.5cm}\includegraphics[scale=1.0]{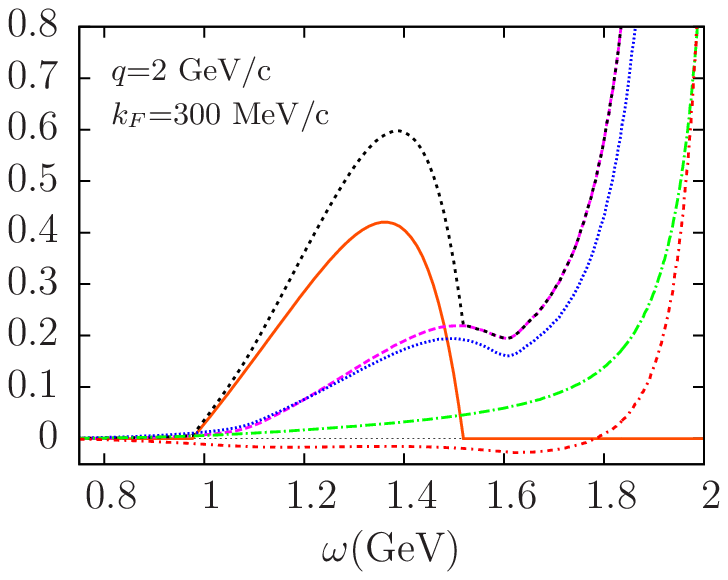}
\caption{(Color online) The 2p-2h MEC transverse response $R^T_{\rm MEC}$ and the separate $\Delta\Delta$, $\pi\pi$ and $\pi\Delta$-interference components
plotted versus $\omega$. The free RFG transverse response (red curves) is also shown for reference.}
\label{fig:fig5}
\end{center}
\end{figure}
%

\begin{figure}[!htb]
\begin{center}
  \includegraphics[scale=1.0]{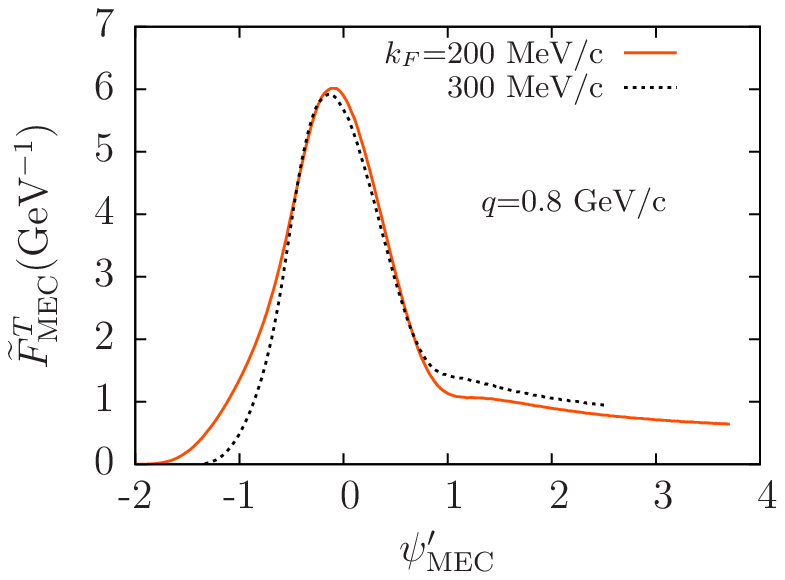}\hspace{0.5cm}\includegraphics[scale=1.0]{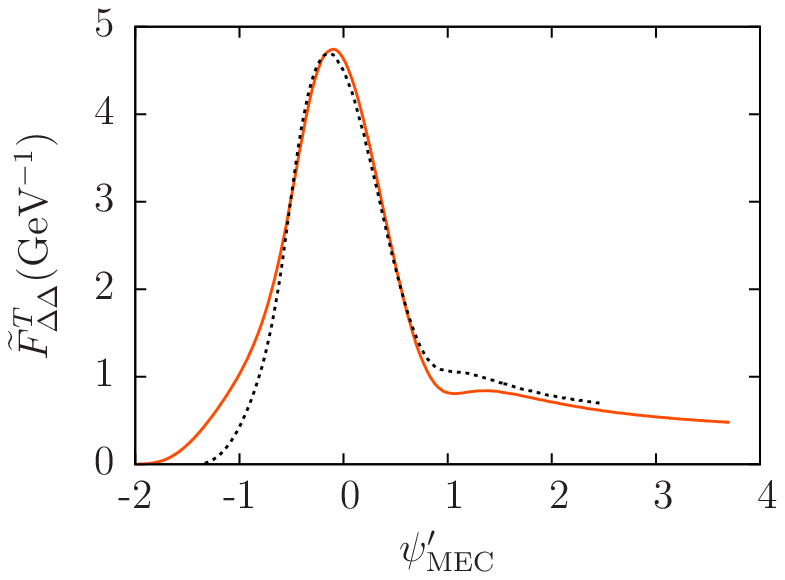}\\
  \includegraphics[scale=1.0]{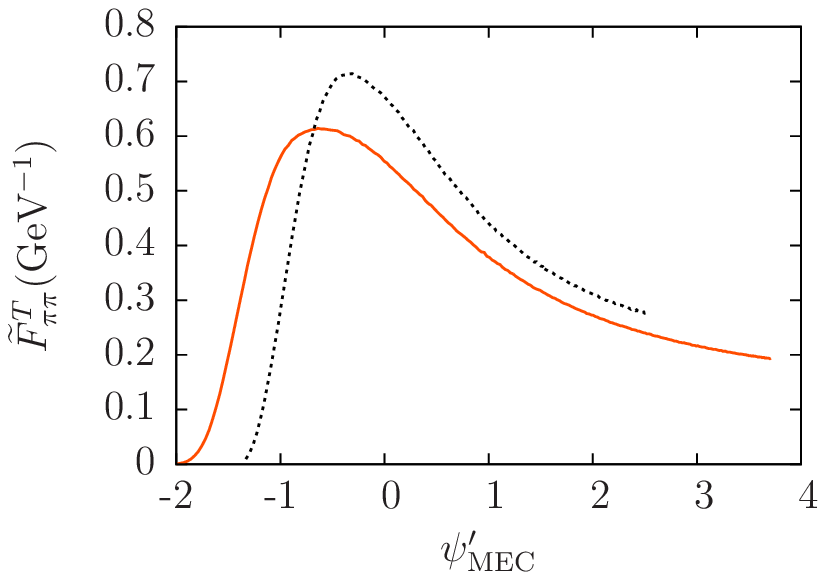}\hspace{0.5cm}\includegraphics[scale=1.0]{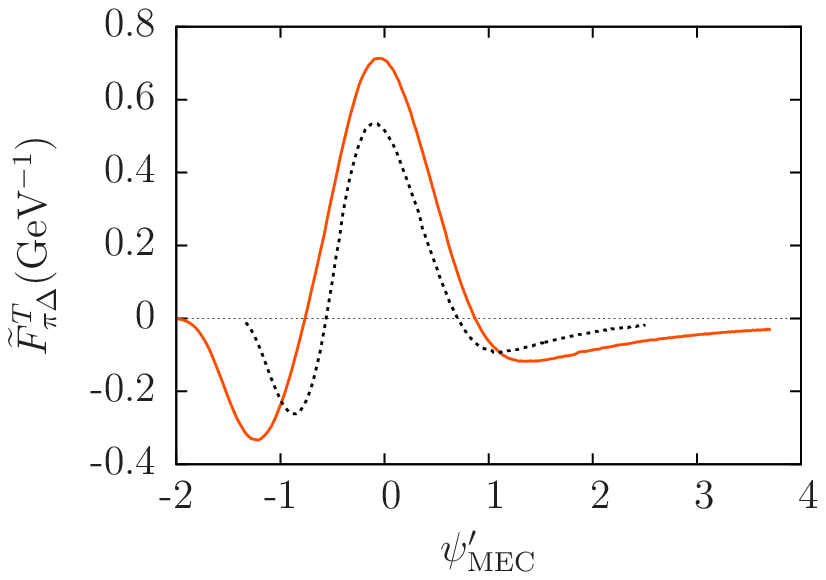}
\caption{(Color online) The scaled 2p-2h MEC response $\widetilde F^T_{\rm MEC}$ defined in Eq. (\ref{eq:tilf}) and the separate $\Delta\Delta$, $\pi\pi$ and $\pi\Delta$-interference components plotted versus $\psi^\prime_{\rm MEC}$  for $q$=800 MeV/c and $k_F$=200 and 300 MeV/c.}
\label{fig:fig6}
\end{center}
\end{figure}
%

\begin{figure}[!htb]
\begin{center}
  \includegraphics[scale=1.0]{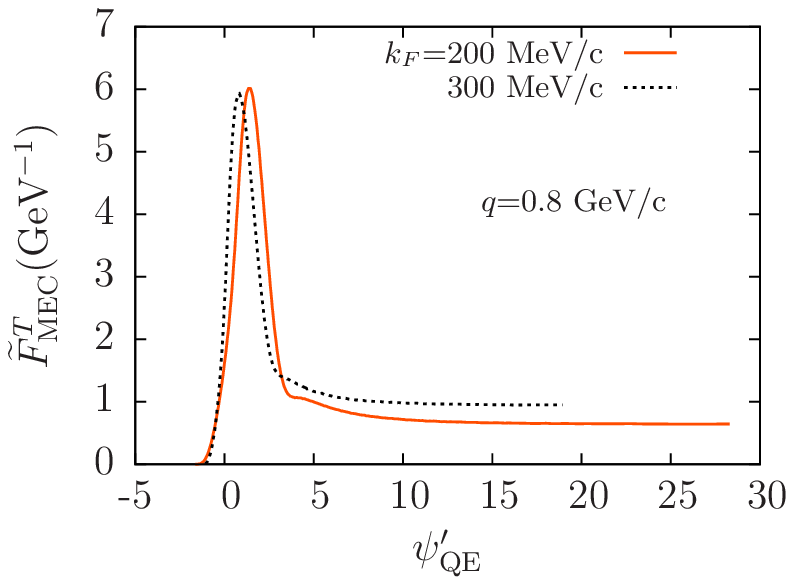}\hspace{0.5cm}\includegraphics[scale=1.0]{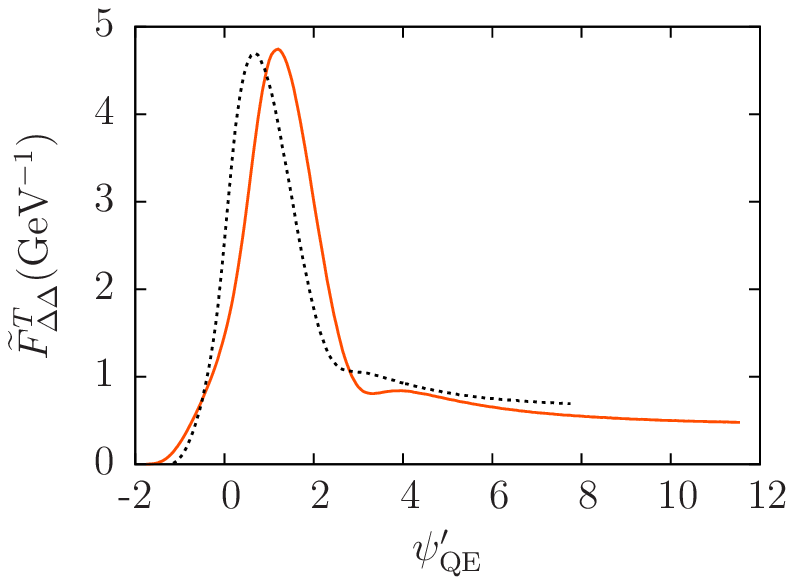}\\
  \includegraphics[scale=1.0]{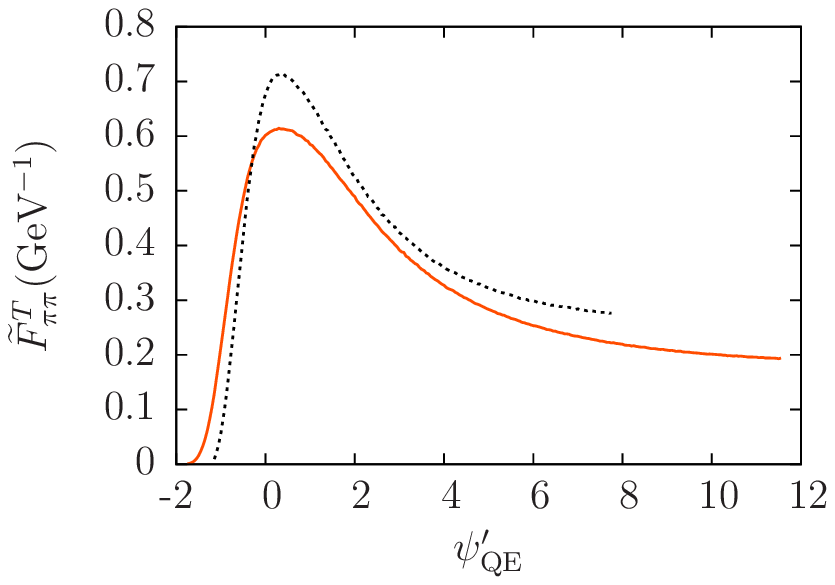}\hspace{0.5cm}\includegraphics[scale=1.0]{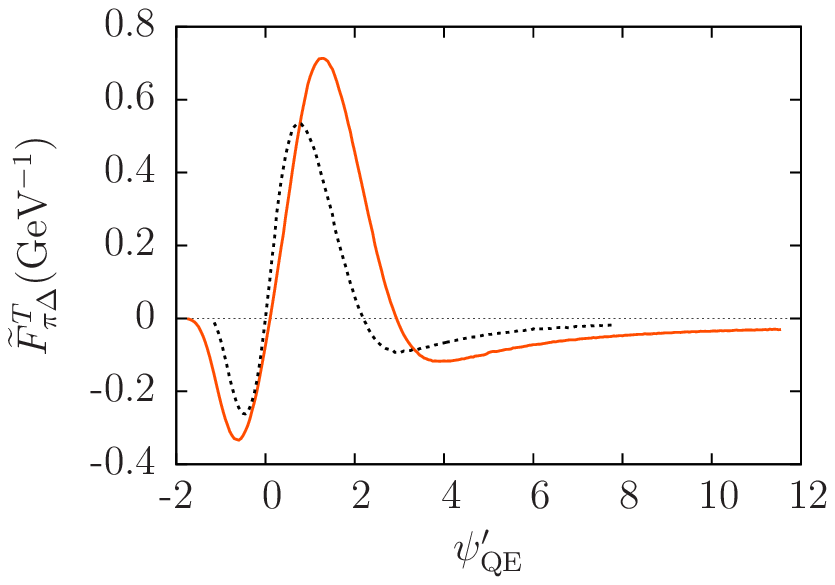}
\caption{(Color online) The scaled 2p-2h MEC response $\widetilde F^T_{\rm MEC}$ defined in Eq. (\ref{eq:tilf}) and the separate $\Delta\Delta$, $\pi\pi$ and $\pi\Delta$-interference components plotted versus $\psi^\prime_{\rm QE}$ for $q$=800 MeV/c and $k_F$=200 and 300 MeV/c.}
\label{fig:fig7}
\end{center}
\end{figure}

%
\begin{figure}[!htb]
  \begin{center}
    \includegraphics[scale=0.65]{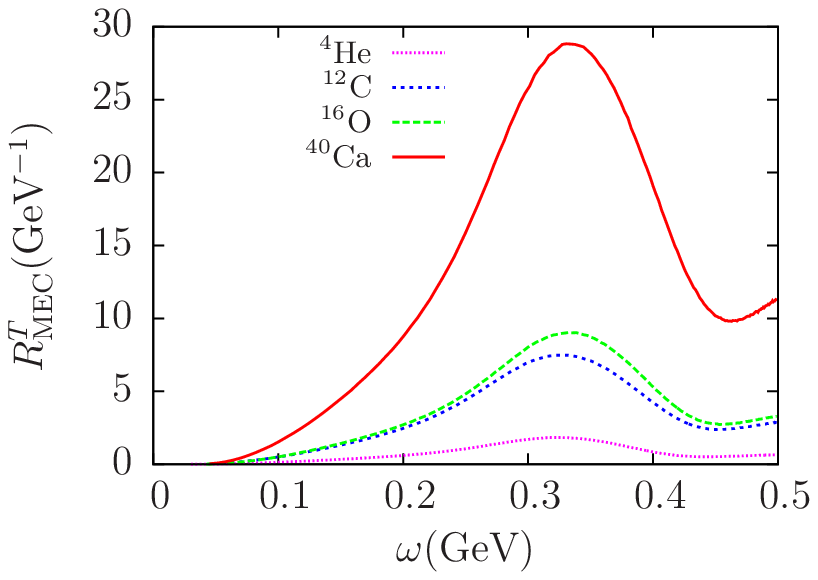}\hspace{0.5cm}\includegraphics[scale=0.65]{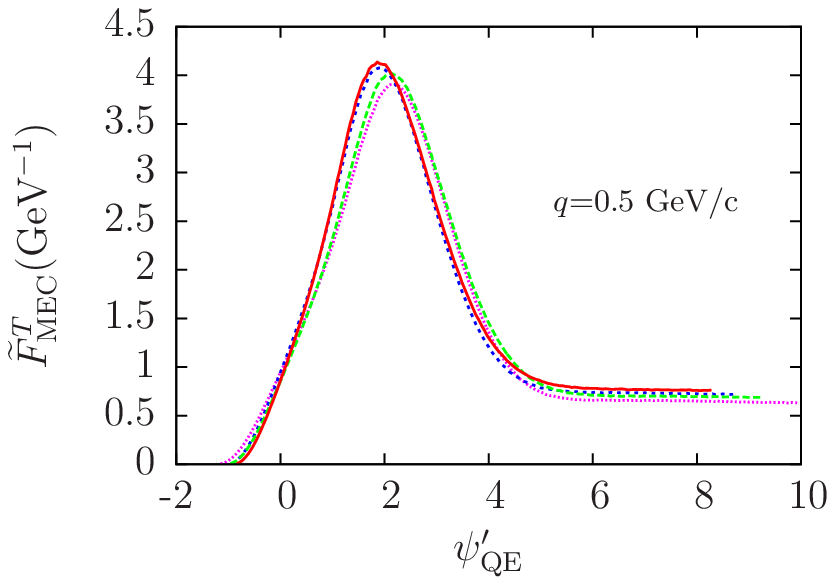}\hspace{0.5cm}\includegraphics[scale=0.65]{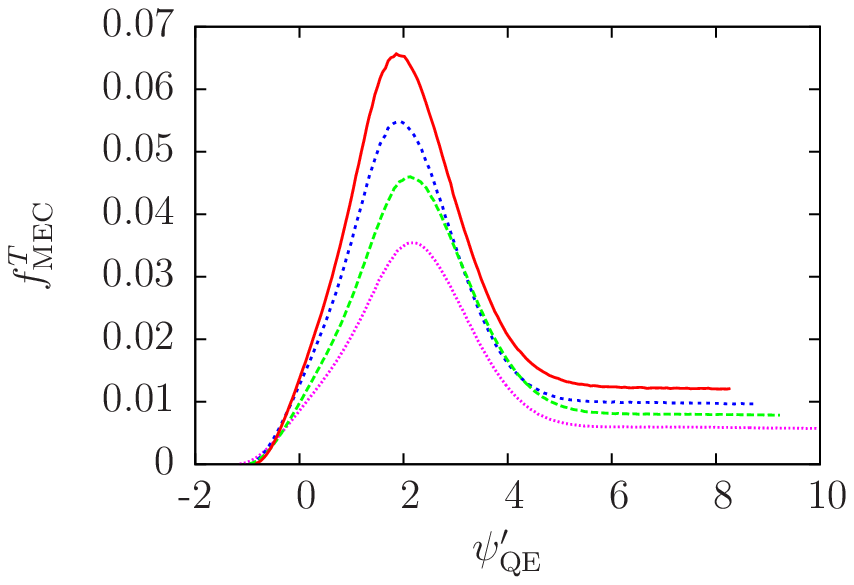}
    \\
    \includegraphics[scale=0.65]{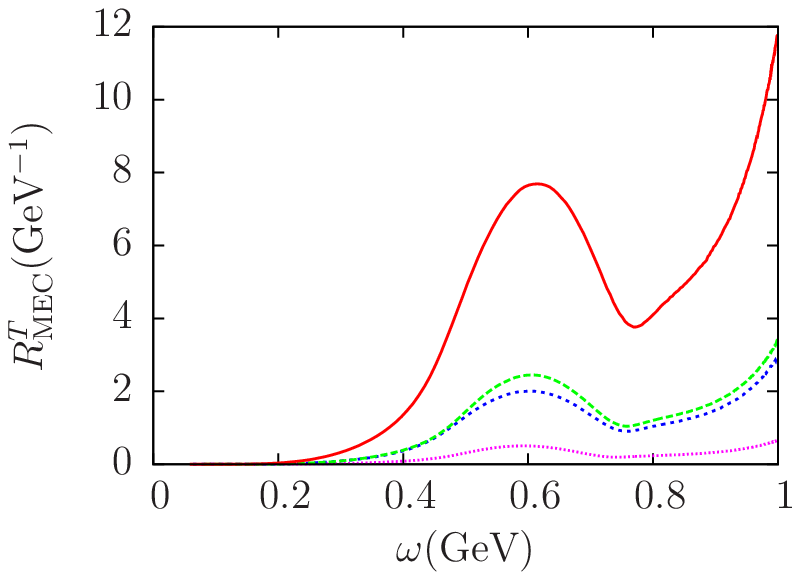}\hspace{0.5cm}\includegraphics[scale=0.65]{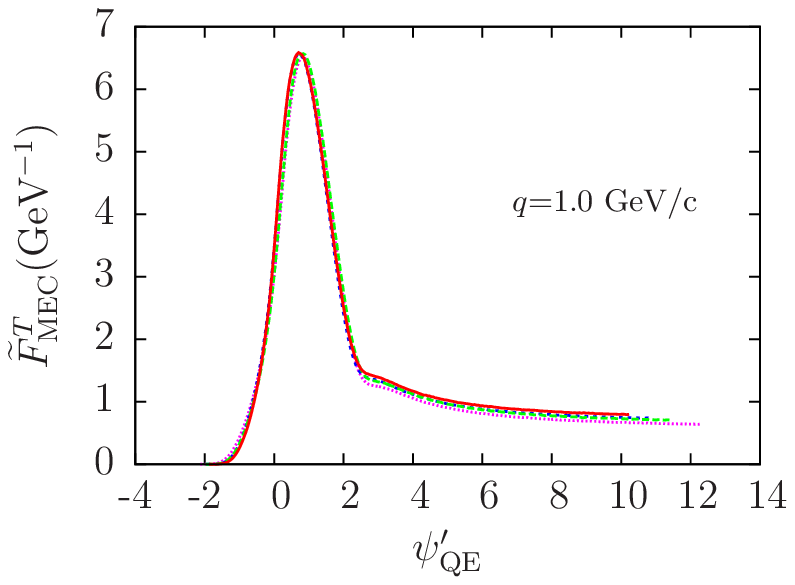}\hspace{0.5cm}\includegraphics[scale=0.65]{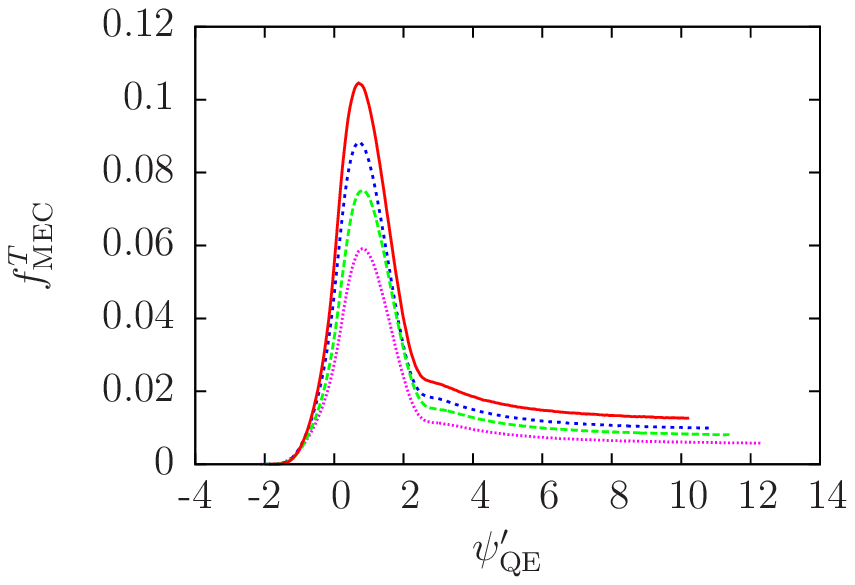}
\\
    \includegraphics[scale=0.65]{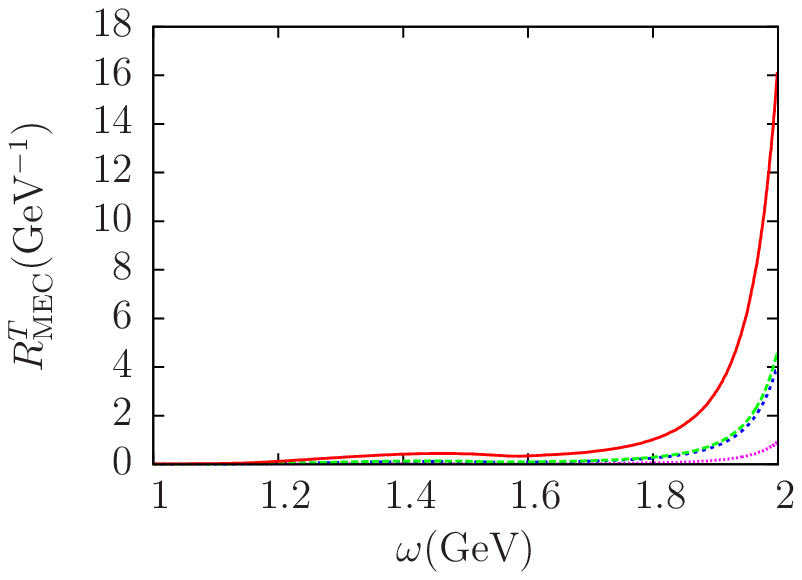}\hspace{0.5cm}\includegraphics[scale=0.65]{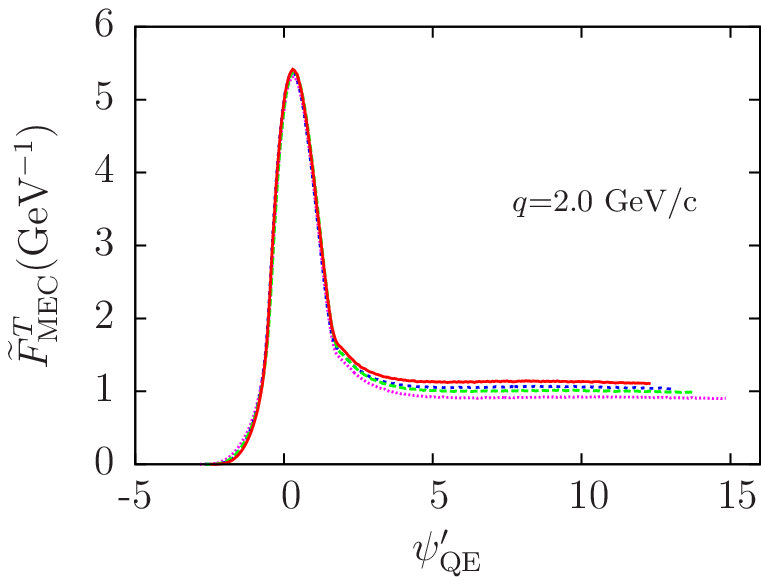}\hspace{0.5cm}\includegraphics[scale=0.65]{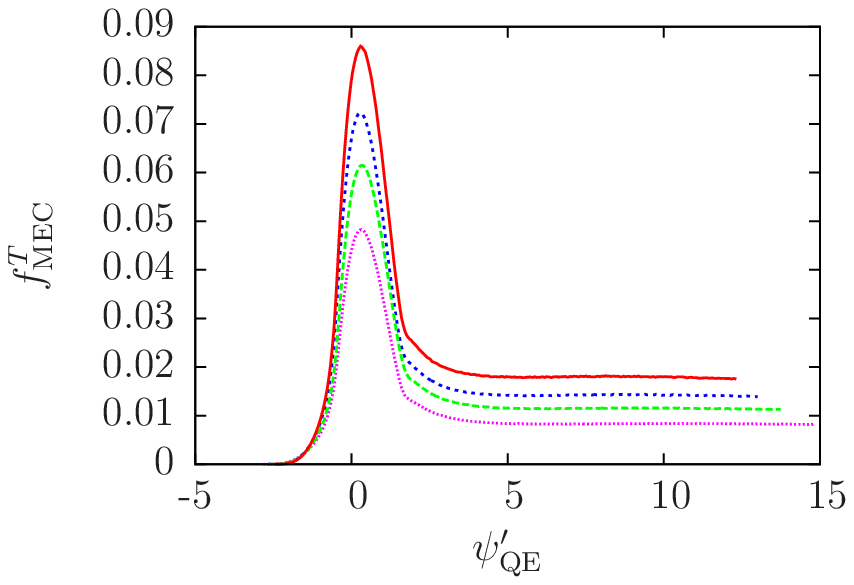}
    \caption{(Color online) The 2p-2h MEC response (first column), the corresponding scaled response $\widetilde F^T_{\rm MEC}$ defined by Eq. (\ref{eq:tilf}) (second column) and the superscaling function defined by  
  Eq. (\ref{eq:fsup}) (third column) for four nuclei and three values of momentum transfer $q$.}
\label{fig:fig8}
\end{center}
\end{figure}

 In the results shown here we take $Z=N$ and we use the Hoeler
parametrization for the proton and neutron magnetic form factors. The
case of asymmetric nuclei, $Z\neq N$ requires more involved
formalism and will be addressed in future work, although preliminary
studies indicate that the qualitative behavior with $k_F$ does not
change dramatically unless $N-Z$ is very large. In particular, the present study can yield
valuable information on how to extrapolate the results obtained for
scattering on $^{12}$C not only to $^{16}$O but also to $^{40}$Ar, a
nucleus widely used in ongoing and future neutrino experiments.

In Fig. \ref{fig:fig1} we display $R^T_{\rm MEC}$ as a function of the
energy transfer $\omega$ for momentum transfers $q$ ranging from 50 to
2000 MeV/c and three values of the Fermi momentum $k_F$
from 200 to 300 MeV/c. 

To illustrate the $k_F$-behavior of the response, we now fix the
momentum transfer to a specific value. In the upper panels of Fig. \ref{fig:fig2} we
show the response $R^T_{\rm MEC}$ and the reduced response $F^T_{\rm
  MEC}$ for $q$=800 MeV/c and the same three values of $k_F$ used above. It clearly
appears that the 2p-2h response, unlike the 1-body quasielastic one,
increases as the Fermi momentum increases.   In the lower panels of Fig. \ref{fig:fig2} we display the scaled 2p-2h
MEC response, defined as
\be \widetilde F^T_{\rm
  MEC}\left(\psi^\prime_{\rm MEC}\right) \equiv \frac{F^T_{\rm
    MEC}}{\eta_F^2} \,,
\label{eq:tilf}
\ee namely the reduced response divided by $\eta_F^2 \equiv
(k_F/m_N)^2$, as a function of the MEC scaling variable
$\psi^\prime_{\rm MEC}(q,\omega,k_F)$ (left panel) and of the
quasielatic one $\psi^\prime_{\rm QE}(q,\omega,k_F)$ (right panel).
The MEC scaling variable is defined in the Appendix, in analogy with the usual QE scaling variable \cite{scaling}.
  The results
show that the reduced 2p-2h response roughly scales as $k_F^2$ when
represented as a function of $\psi^\prime_{\rm MEC}$ (Fig. 2c), {\it i.e.}, the scaled 2p-2h MEC response shown there coalesces at the peak into a universal result. This
scaling law is very accurate at the peak of the 2p-2h response, while
it is violated to some extent at large negative values of the scaling
variable. Fig. 2d shows that in this ``deep scaling'' region it is more appropriate to
use the usual scaling variable $\psi^\prime_{\rm QE}$ devised for
quasielastic scattering.
  This latter region was previously investigated in \cite{DePace04}: in that study the specific cases of $^{12}$C ($k_F$=228 MeV/c) and $^{197}$Au ($k_F$=245 MeV/c) were considered and the superscaling functions $f$ were plotted versus $\psi^\prime$ ($f^T_{\rm MEC}$ and $\psi^\prime_{\rm QE}$ in the present work) together with JLab data at electron energy $\epsilon$=4.045 GeV and scattering angle $\theta$=23$^0$ and 30$^0$ -- see Fig. 7 in \cite{DePace04}.
  Following \cite{DePace04} $f^T_{\rm MEC}$ is defined by
  \be f^T_{\rm  MEC} \equiv F^T_{\rm MEC} \times k_F \,.
\label{eq:fsup}
\ee
  There one observes two things: (1) the usual scaling, {\it i.e.}, {\em not} the scaling behavior found in the present study at the peak of the MEC response, is reasonably compatible with the spread found in the data, 
and (2) at very high momentum transfers the 2p-2h MEC contributions are very significant in this deep scaling region, to the extent that they may even provide the dominant effect.

For completeness, in Fig. \ref{fig:fig3} we show results at $q$=2 GeV/c using the two types of $k_F$-scaling behavior.
In particular, in the right-hand panel where the usual superscaling results are presented  it should be emphasized that, for the most negative values of $\psi^\prime_{\rm QE}$ (the deep scaling region), the data analyzed in \cite{DePace04} fall well inside the range spanned by the upper curve ($k_F$=200 MeV/c) and the middle curve ($k_F$=250 MeV/c). 

In Fig. \ref{fig:fig4} the scaled 2p-2h MEC response is now plotted versus $\psi^\prime_{\rm MEC}$ for four values of $q$. Here we see 
that the same $k_F$-dependence is valid for different values of $q$ as long as Pauli blocking is not active, namely $q>2 k_F$. At lower $q$ and in the deep scaling region this type of scaling is seen to be broken (see also above).

A closer inspection of the scaling properties of the 2p-2h response is
presented in Figs. \ref{fig:fig5} -- \ref{fig:fig7}. In Fig. \ref{fig:fig5} the separate contributions of
$\Delta\Delta$, $\pi\pi$ and $\pi\Delta$-interference terms
are displayed for two values of $q$ and two values of $k_F$.  In 
Fig. \ref{fig:fig6}
the corresponding scaled responses are displayed as
functions of the variable $\psi^\prime_{\rm MEC}$: it appears that all
contributions roughly grow as $k_F^2$, the quality of scaling being
better for the $\Delta\Delta$ piece than for the other two
contributions. It is interesting to observe that at high momentum
transfer the total MEC response scales better than the pure $\Delta$
piece around the peak, indicating a compensation of scaling violations
between the three terms.  We notice that scaling violations
are more sizeable away from the peak: in 
Fig. \ref{fig:fig7}
it is shown that in this region the quasielastic scaling variable, which
appears to be more suitable to describe the pure pionic ($\pi\pi$)
and interference ($\pi\Delta$) terms, gives a better scaling of
second kind.

Finally, focusing on practical cases, in Fig. \ref{fig:fig8} we show $R^T_{\rm MEC}$ versus $\omega$, together with $\widetilde F^T_{\rm MEC}$ anf $f^T_{\rm MEC}$ versus $\psi^\prime_{\rm qe}$ for three values of $q$ and for the symmetric nuclei $^{4}$He, $^{12}$C, $^{16}$O and $^{40}$Ca. The cases of $^{12}$C and $^{16}$O are clearly relevant for ongoing neutrino oscillation studies, whereas the case of $^{40}$Ca is a symmetric nucleus lying close to the important case of $^{40}$Ar.
For comparison, $^{4}$He is also displayed and, despite its small mass, is seen to be ``typical''. In contrast, the case of $^{2}$H, whose Fermi momentum is unusually small ($k_F$= 55 MeV/c), was also explored and found to be completely anomalous: the MEC responses ($R^T_{\rm MEC}$) and superscaling results ($f^T_{\rm MEC}$) were both too small to show in the figure.

 \section{Conclusions}

Summarizing, we have shown that the 2p-2h MEC response
function per nucleon roughly grows as $k_F^2$ for Fermi momenta
varying from 200 to 300 MeV/c.  This scaling law is excellent around
the MEC peak for high values of $q$, it starts to break down around $q = 2
k_F$, and gets worse and worse as $q$ decreases.  This behavior must
be compared with that of the 1-body response, which scales as $1/k_F$:
hence the relative importance of the 2p-2h contribution grows as
$k_F^3$.  This result allows one to get an estimate of the relevance
of these contributions for a variety of nuclei, of interest in ongoing
and future neutrino scattering experiments, and should facilitate the
implementation of 2p-2h effects in Monte Carlo generators.
Finally, in the deep scaling region the MEC response is found to be significant and to scale not as $k_F^2$, but rather more as $1/k_F$.

\section*{Appendix}
The MEC scaling variable is defined as  
\be \psi^\prime_{\rm MEC}
(q,\omega,k_F) \equiv \frac{1}{\sqrt{\xi_F^{eff}(q)}} \,
\frac{\lambda^\prime_{\rm MEC} - \tau^\prime_{\rm MEC}
  \rho^\prime_{\rm MEC}}{\sqrt{ (1+\lambda^\prime_{\rm MEC}
    \rho^\prime_{\rm MEC})\tau^\prime_{\rm MEC} + \kappa
    \sqrt{\tau^\prime_{\rm MEC} \left(1+\tau^\prime_{\rm MEC}
      \rho^{\prime \ 2}_{\rm MEC}\right)} }} \,, 
\ee 
where 
\be
\lambda^\prime_{\rm MEC}\equiv\frac{\omega^\prime_{\rm MEC}}{2m_N} \,,\ \ \ \ 
\kappa \equiv\frac{q}{2m_N} \,,\ \ \ \ 
\tau^\prime_{\rm MEC}\equiv \kappa^2-(\lambda^\prime_{\rm MEC})^2 \,,\ \ \ \ 
\omega^\prime_{\rm MEC} \equiv \omega-E^{shift}_{\rm MEC}(q) \,,
\ \ \ \ 
\rho^\prime_{\rm MEC} \equiv 1 + \frac{1}{4\tau^\prime_{\rm MEC}} \left(\frac{m_*^2}{m_N^2}-1\right) \,.
\ee
  \begin{table}[ht]
  \centering
  \label{tab:table1}
\fbox{
  \begin{tabular}{c|c|c|c|c|c|c}
    $m_*$(MeV/c$^2$) \ & \ $\alpha$\  & \ $\beta$\ &\ $\gamma$\ &\ $E_0$(MeV) \ & \ $E_1$(MeV)\ &\ $E_2$(MeV) \\
    \hline
    1170 &\  1.3345\ &\ 30.73\ &\ 0.85\ & 42.718 & -70.0 & 37.0 \\
  \end{tabular}
}
\caption{the parameters entering the definition of $\psi^\prime_{\rm MEC}$ for $^{12}$C.}
\end{table}

The functions 
\be
\xi_F^{eff}(q)=\sqrt{1+\left[\alpha \left(1+\beta e^{-w\gamma}\right)\eta_F\right]^2}-1\ee and \be E^{shift}_{\rm MEC}(q)=E_0+E_1 t+E_2 t^2 \,,\ee with $w=q/1000$ and $t=(q-500)/1000$  with $q$ in MeV/c 
are chosen in such a way that the maxima of the 2p-2h response at different values of $q$ align at $\psi^\prime_{\rm MEC}=0$.  The values of the parameters for the case of $^{12}C$ are given in Table 1; the same values are used for all the choices of $k_F$ and the results shown in Fig. \ref{fig:fig4} indicate that this procedure is successful.

The usual
definition of $\psi^\prime_{\rm QE}$ can be recovered from the above
equations by setting $m_*=m_N$ (hence $\rho^\prime=1$).

\section{Acknowledgements}

This work has been partially supported by the INFN under project
MANYBODY, by the Spanish Ministerio de Economia y Competitividad and ERDF (European Regional Development Fund) under contracts FIS2014-59386-P, FIS2014-53448-C2-1, by the Junta de Andalucia (grants No. FQM-225, FQM160), and part (TWD) by the U.S. Department of Energy under cooperative agreement DE-FC02-94ER40818. IRS acknowledges support from a Juan de la Cierva fellowship from MINECO (Spain). GDM acknowledges support from a Junta de Andalucia fellowship (FQM7632, Proyectos de Excelencia 2011).

\end{document}